# Characteristics and Large Bulk Density of the C-type Main-Belt Triple Asteroid (93) Minerva

F. Marchis[1,2], F. Vachier[2], J. Ďurech[3], J.E. Enriquez[1], A. W. Harris[4], P. A. Dalba[1,5], J. Berthier[2], J. P. Emery[6], H. Bouy[7], J. Melbourne[8], A. Stockton[9], C. D. Fassnacht[10], T. J. Dupuy[9], J. Strajnic[11]

1. Carl Sagan Center at the SETI Institute, 189 Bernardo Av, Mountain View CA 94043, USA
2. Institut de Mécanique Céleste et de Calcul des Éphémérides, Observatoire de Paris, UMR8028 CNRS, 77 av. Denfert-Rochereau 75014 Paris, France
3. Astronomical Institute, Faculty of Mathematics and Physics, Charles University in Prague, V Holesovickach 2, 18000 Prague, Czech Republic
4. DLR Institute of Planetary Research, Rutherfordstrasse 2, 12489 Berlin, Germany

5. University of California at Berkeley, Department of Astronomy, B-20 Hearst Field Annex #3411, Berkeley CA 94720, USA
6. Earth and Planetary Sciences, University of Tennessee, 306 Earth and Planetary Sciences Building, Knoxville, TN 37996-1410, USA
7. Centro de Astrobiología, INTA-CSIC, PO Box 78, 28691, Villanueva de la Cañada, Madrid, Spain
8. California Institute of Technology MS 301-17, Pasadena, CA 91125 USA
9. Institute for Astronomy, University of Hawaii, 2680 Woodlawn Drive, Honolulu, HI 96822, USA
10. Department of Physics, University of California Davis, 1 Shields Avenue, Davis, CA 95616, USA
11. Direction de l'enseignement supérieur et de la recherche, Rectorat d'Aix-Marseille, Place Lucien Paye, 13100 Aix-en-Provence, France

Pages: 50
Tables: 5
Figures: 9
Supplement Materials: 3

*Corresponding author:*
**Franck Marchis**
SETI Institute
Carl Sagan Center
189 Bernardo Av
Mountain View CA 94043
USA
fmarchis@seti.org
Phone: +1 650-810-0236

**Abstract:**
From a set of adaptive optics (AO) observations collected with the W.M. Keck telescope between August and September 2009, we derived the orbital parameters of the most recently discovered satellites of the large C-type asteroid (93) Minerva. The satellites of Minerva, which are approximately 3 and 4 km in diameter, orbit very close to the primary (~5 & ~8 × $R_p$ and ~1% & ~2% × $R_{Hill}$) in a circular manner, sharing common characteristics with most of the triple asteroid systems in the main-belt. Combining these AO observations with lightcurve data collected since 1980 and two stellar occultations in 2010 & 2011, we removed the ambiguity of the pole solution of Minerva's primary and showed that it has an almost regular shape with an equivalent diameter $D_{eq}$ = 154 ± 6 km in agreement with IRAS observations. The surprisingly high bulk density of 1.75 ± 0.30 g/cm³ for this C-type asteroid, suggests that this taxonomic class is composed of asteroids with different compositions, For instance, Minerva could be made of the same material as dry CR, CO, and CV meteorites. We discuss possible scenarios on the origin of the system and conclude that future observations may shine light on the nature and composition of this fifth known triple main-belt asteroid.

# 1 Introduction

The serendipitous discovery of Dactyl, a companion of the main-belt asteroid (243) Ida, seen during the Galileo spacecraft voyage to Jupiter (Chapman et al. 1995), gave birth to a new field of study for small solar system bodies. Today, using techniques such as high angular resolution imaging with the Hubble Space Telescope (HST), adaptive optics (AO) on ground-based telescopes, radar observations and photometric studies, about 200 multiple asteroids (asteroids with one or several moons) are known in all populations of small solar system bodies, from the Near-Earth Asteroids to the Kuiper Belt Objects. The study of these multiple asteroids is an opportunity for planetary astronomers to obtain insights on the asteroids such as their masses and densities and how these quantities possibly relate to their compositions, see e.g. Marchis et al. 2012a. Their existence and the understanding of their formation and evolution also provide a direct window to the history of our solar system.

In this work, we describe a study of one of these recently discovered triple asteroid systems. (93) Minerva is a large ($H_v$ = 7.7, $P_{spin}$ = 5.97909 hours in Tungalag 2002) asteroid discovered by J.C. Watson at Ann Arbor, MI, USA in 1867. Located in the middle of the main-belt ($a_{Minerva}$ = 2.75 AU, $e_{Minerva}$ = 0.14, $i_{Minerva}$ = 9 deg), it was initially classified as a member of an old collisional family (1.2 ± 0.4 Gyrs) named "Gefion" consisting of 973 members based on Nesvorny et al. (2005). Because most of the members of this family are identified as S-type, and (93) Minerva is known to be a $C_b$-type (Lazarro et al. 2004) or C-type (deMeo et al. 2009), the large Minerva is likely to be an interloper in the Gefion family.

The size and geometrical albedo of Minerva were reported in several mid-infrared surveys such as IRAS ($D$ = 142 ± 4 km, $p_v$ = 0.073 ± 0.004 in Tedesco et al. 2002), MSX ($D$ = 157 ± 3 km, $p_v$ = 0.060 ± 0.002 in Price et al. 2001), AKARI ($D$ = 147 ± 2 km, $p_v$ = 0.068 ± 0.003 in Usui et al.

2011) and in a reanalysis of IRAS & MSX data ($D$ = 165 ± 8 km, $p_v$ = 0.0597 ± 0.0021 in Ryan & Woodward, 2010). There is a clear discrepancy in the sizes of Minerva derived from mid-infrared catalogs (up to ~23 km), indicative of different viewing geometries of the asteroid at the time of the observations, but also linked to a slightly different modeling in these works. The disparate diameters could be due to the use of different absolute magnitudes ($H_v$) for Minerva, assumptions on the thermal model and distinctive wavelength bands of observations between these surveys.

In the next section of this work, we describe observations of (93) Minerva from different sources. The key finding of this work is the discovery of two satellites around the large asteroid, which was made possible by W.M. Keck AO observations. We also combined our analysis with a large set of lightcurve data and stellar occultation data already published in the literature. Section 3 describes how a coherent analysis of these disparate sets of data allowed us to constrain the pole orientation, size and shape of the primary of the system. The mutual orbits of the satellites are determined in Section 4 and compared to other known triple asteroid systems. We discuss the composition and the origin of the system based on the first direct measurement of its density in Section 5. In Section 6 we discuss the origin of the system based on the orbital parameters and contrast our study with other multiple asteroids. A summary of our findings and future work is discussed in Section 7.

## 2   Observations and data processing

### 2.1   Adaptive Optics Observations

Observations of (93) Minerva were collected on August 16, 2009 using the W.M. Keck II telescope located atop Mauna Kea on the Big Island of Hawaii. The Nasmyth platform of this 10m telescope has been host to the NIRC2 infrared camera equipped with an AO system since 2001 (Wizinowich et al. 2000; van Dam et al. 2004). Thanks to the large aperture size of the telescope and the potential of AO, the angular resolution expected on these images for a V~12 bright target (see **Table 1**) is ~40-50 milli-arcsec (*mas*), very close to the diffraction limit of the telescope in FeII ($\lambda_c$ = 1.6455 µm, $\Delta\lambda$=0.0256 µm) and Kcont ($\lambda_c$ = 2.2706 µm, $\Delta\lambda$=0.0296 µm). The first images were taken in these broadband filters (**Table 1**) from 13:38 to 13:45 UT using the narrow camera with a pixel scale of 9.94 *mas*. A final image was obtained using our automatic pipeline while observing at the telescope by shift-adding 3 to 6 frames with an exposure time of 60 seconds (30 seconds x 2 coadds). These frames were flat-field corrected, and we used a bad-pixel suppressing algorithm to improve the quality before shift-adding them. After this basic data processing, the final images revealed the presence of one small satellite at 3 o'clock at 0.42 arcsec (see **Fig. 1**). Additional observations taken at 13:56 UT revealed the existence of an even closer (~0.26 arcsec) companion at 5 o'clock on the image. These existence of these satellites was corroborated by additional data taken one hour later (after 14:40 UT), which confirmed that these companions are gravitationally linked to the primary and orbit around the primary in the clockwise direction.
**[INSERT HERE FIG. 1 & TABLE 1]**

Additional data were collected from September 6 to September 28, 2009 when the asteroid was approaching its opposition (elongation reached 160 deg on September 28). Because the asteroid was brighter (V ~ 11.7) and closer (distance to Earth $d$ ~ 1.80 AU), the AO correction was significantly better and the two satellites were easily detectable in the images after the basic processing described above. (**Fig. 2**). These additional observation runs confirmed the genuineness of our discovery and allowed us to derive the characteristics of this multiple asteroid system.
**[INSERT HERE FIG. 2]**

(93) Minerva is the fifth triple asteroid discovered among the main belt asteroids (Marchis et al. 2009), after 87 Sylvia (Marchis et al., 2005), 45 Eugenia (Marchis et al., 2007), 3749 Balam (Marchis et al., 2008c) and (216) Kleopatra (Marchis et al., 2008d). For the sake of clarity we adopt from now on the designation *(93) Minerva* for the whole system and *Minerva* for the primary alone. Minerva's companions are 5-6 magnitudes fainter than the primary body and are separated by about 0.26 and 0.42 arcsec respectively (**Fig. 1**). They are the smallest and closest satellites of a large asteroid ever seen. This discovery illustrates the improvement in image quality and sensitivity achieved with the AO technology over the past 15 years (Marchis et al. 2012b).

The size of the satellites can be estimated by assuming that both the moons and the primary have the same albedo and comparing their flux ratios ($\phi$). In **Table 2**, we list the effective diameters of satellites ($D_{eff}(satellite)$), derived in each image with the equation $D_{eff}(satellite) = \phi^{1/2} \times D_{eff}(primary)$ where $D_{eff}(primary)$ is the measured effective diameter on the resolved adaptive optics observations. For Minerva I, the outer moon, and Minerva II, the inner moon, we derived the following diameters by averaging $D_{eff}(satellite)$ measurements: $D_I = 3.6 \pm 1.0$ km and $D_{II} = 3.2 \pm 0.9$ km.
**[INSERT HERE TABLE 2]**

A single star (BD+11 229), with a brightness similar to (93) Minerva (*V*=11.06) and located at ~0.4 deg from the asteroid, was observed shortly after the observation of (93) Minerva. This additional set of observations is useful to estimate the quality of the data, check for possible artefacts in the Point Spread Function (PSF) of the instrument and deconvolve the data during further analysis. This PSF image was observed at 13:50 UT using an FeII filter with the narrow camera and a total exposure time of 25 seconds per frame (5 seconds x 5 coadds) and was processed in a manner similar to the asteroid observations. We estimated the angular resolution on the Minerva data (~41 *mas*) by measuring the full width at half maximum (FWHM) of the PSF. Interestingly, the FWHM measurement of the Minerva observations varies from 0.10 to 0.12 arcsec, implying that the primary is resolved with 2-3 elements of resolution.

Because the final images have a high signal-to-noise ratio varying from 1400 to 2800, and because it is known that the Keck AO correction is relatively stable for bright V~11-12 targets, we can apply the AIDA myopic deconvolution algorithm (Hom et al. 2005) to the observations using the PSF frames recorded on August 16 as an input estimate of the real PSF. The deconvolved frames shown in **Figures 1 & 2** suggest that the projection of Minerva's primary is almost spherical with a maximum ratio *2a'/2b'* (ratio of the projected major (*a'*) and minor (*b'*) axes derived by fitting an ellipse to the resolved primary image) of 1.2 (*2a'/2b'* = 1.1 in average).

From these deconvolved images, we extracted the silhouette of the Minerva primary, which will be used to derive the shape of Minerva in Section 4. From simulations of asteroid observations, Marchis et al. (2012b) showed that the typical error on the major-axis estimate of an asteroid resolved with ~3 elements of resolution (SNR~2,000, FWHM(PSF) ~ 40 mas) is 3%, which corresponds to 4 *mas* or 6 km at the distance of Minerva for the epochs of observations.

## 2.2 Lightcurve data

We gathered photometric lightcurve observations of (93) Minerva from various sources listed in **Table 3** from 1980 to 2012. The first set of lightcurves taken from 1980 to 2009 were published in Harris and Young (1999), Debehogne et al. (1982), Harris and Young (1999), Denchev (2000) and Torppa et al. (2008). We retrieved them from the Asteroid Photometric Catalog[1], a repository web site for asteroid lightcurves. Starting in 2011, the robotic 0.6m super-LOTIS telescope located at Kitt Peak was also used to record R band observations. The data were reduced using a classic relative photometry method consisting in comparing the flux of the asteroid with the flux of nearby stars over the course of the night. The last night of the super-LOTIS data (February 18 2012) was affected by poor weather conditions. On February 4 2011, the 0.8m telescope of the Observatoire de Haute-Provence collected photometric observations of over two-thirds of the spin period.

**[INSERT HERE TABLE 3]**

The lightcurves in **Fig. 3** display relative intensity versus time. They have a remarkably low amplitude (less than 0.25 mag) even when the asteroid was observed edge-on in February and March 2011, suggesting that the asteroid primary, which dominates the flux variation in the lightcurves, is almost spherical. Based on the flux ratio between the moons ($D_s$ ~ 3 km) and the primary ($D_p$ ~ 141 km by Tedesco et al. 2002) we expect a mutual event to generate a small attenuation (~0.001 mag) on the lightcurve. Because this signature of mutual event is significantly smaller than the accuracy on our lightcurve, none were reported in the lightcurve sample.

**[INSERT HERE FIG 3]**

## 2.3 Stellar Occultation data

Similarly to AO images, a successful observation of an occultation constrains the size of the asteroid, its spin axis orientation, and its shape (Ďurech et al. 2011). There are ten occultations by Minerva reported in Dunhan et al. (2012) but only three events have more than two positive chords measured. The events took place on 1982-11-22 (9 chords, max duration ~11s), 2010-12-24 (9 chords, max duration ~12s) and 2011-01-28 (4 chords, max duration~12s). In this work, we used only occultations from 2010 and 2011, since the 1982 event was observed visually by all observers and Ďurech et al. (2011) showed that there are systematic shifts in time, likely due to the reaction times of the observers.

**Figure 4** shows the observed chords of the stellar occultation. The solid chords were observed using video or CCD, the one dashed chord was observed visually and the dotted line in the first occultation is the negative observation with no occultation detected. These data will be used to constrain the size and shape of Minerva's primary in Section 4.1.

---

[1] http://asteroid.astro.helsinki.fi/apc/asteroids/93

Unfortunately, no observers reported the detection of secondary stellar occultation events due to the satellites. Such a detection would have allowed the direct measure the sizes of the satellites as Descamps et al. (2008) reported for Linus, companion of (22) Kalliope. It would have also refined the orbital parameters of the satellites as shown for Linus and Kalliope in Vachier et al. (2012).

## 3  Pole Solution, Size and Shape of the Primary
### 3.1  Combining LC, AO and Occultation Data

By combining all available lightcurves, AO images and two of the stellar occultations, we derived a unique shape and spin model of Minerva. We used the KOALA code that simultaneously fits the lightcurves, profiles of the AO images and occultation chords to produce a best-fit shape model that is represented by a non-convex polyhedron. The mathematical background of the algorithm was described by Kaasalainen (2011) and the reliability of the method was shown by Marchis et al. (2006) by comparing adaptive optics observations with shape model of several main-belt asteroids.

Before running the KOALA code, we used a lightcurve inversion method (Kaasalainen and Torppa 2001, Kaasalainen et al. 2001) to scan the parameter space of the rotation period and the pole direction. We found a unique solution for the rotation period P = 5.98176 h and two possible solutions for the pole direction: λ = 11 deg, β= +25 deg and λ = 200 deg, β= +22 deg in ecliptic coordinates J2000 (or ECJ2000).

The ambiguity in the pole direction is caused by the proximity of the asteroid orbit to the plane of the ecliptic. However, when the sky-plane projection of the two models was computed for the time of AO observations and the stellar occultations, only the first pole solution agreed. We rejected the second pole solution and started the KOALA code with the initial period and pole of the lightcurve inversion model. The fits to the lightcurves, AO silhouettes and occultations are displayed in **Figures 3, 4 and 5**. The resulting non-convex shape model is shown in **Figure 6**. The rotation period is 5.981767 ± 0.000004 h with a spin pole solution (λ=21 deg, β=21 deg in ECJ2000) with an uncertainty estimated to ± 10 deg.
**[INSERT HERE FIG 4, 5, 6]**

**Figure 4** shows the projected silhouette of the model of Minerva. The overall size and shape of the model are in good agreement with the chords despite some minor discrepancies, which can be explained by timing errors (the reported errors range from 0.01 to 2 seconds) as well as uncertainties in the shape model.

A comparison between the AO profile after deconvolution and the silhouette of the non-convex model is shown in **Figure 5**. The equivalent diameter we derived from these measurements and the stellar occultation observations is $D_{eq}$ = 154 km. The residual mean square error of 3.5 *mas*, corresponds to 6 km on Minerva, less than the 1-sigma error (4 *mas*) that we estimated on the shape profile of Minerva AO images discussed in Section 2.1.

Our pole is the mirror of the adopted orientation by Harris et al. (1989) (λ=230 deg, b=0 deg in ECJ2000), one of the poles of Torppa et al. (2008) (λ=216 deg, β=21 deg in ECJ2000) and the pole solution by Tungalag et al. (2002) (λ=189 deg, β=10 deg in ECJ2000). The addition of the AO and stellar occultation data were indeed crucial to remove the pole ambiguity.

### 3.2   An almost-spherical primary
The non-convex model has a small flattening ($a_0/a_2 \sim 1.25$, with $a_0 > a_1 > a_2$ the main axes of an ellipsoid that fits the shape of Minerva's primary), smaller than the convex model ($a_0/a_2 \sim 1.7$) and the flattening estimate ($a_0/a_2 \sim 1.45$) derived by Harris et al. (1989) that included absolute magnitude measurements in the analysis. The shape model derived by Torppa et al. (2008), which is based on a limited amount of lightcurve data, also suggested that Minerva's primary is regular in shape.

A comparison between the convex model derived from the lightcurves and the absolute magnitude and stellar occultation data collected on December 24, 2010 is displayed in **Figure 4**. This model poorly fits the chords of the stellar occultations whereas the non-convex model gives a better fit. We therefore favor the non-convex shape model with $D_{eq} = 154 \pm 6$ km shown in **Figure 6** for the rest of this work.

The slightly elongated shape of the Minerva primary leads to brightness variations of less than 30% as seen from Earth (e.g. lightcurve observations in March 2011) in agreement with the discrepancy between the mid-infrared size measurements reported in Section 1. In other words, the viewing geometry could explain the differences in the size estimates between IRAS (D = 142 km) and IRAS-MSX (D=165 km) for instance.

### 3.3   Agreement with IRAS observations and absolute magnitude
As discussed in Section 1, several mid-IR estimates of the size of (93) Minerva appear in the literature with a discrepancy of up to 23 km. We checked the validity of the equivalent diameter determined in Section 3.2 by reanalyzing two sightings available in the IRAS catalog (Tedesco et al., 2002) from November 6 1983 at 16:02 and 19:28 UT (i.e. almost half a spin period apart). The fits to the IRAS flux measurements of three different thermal models (see Harris, 1998 and references therein), the standard thermal model (STM), the fast rotating model (FRM), the near-Earth asteroid thermal model (NEATM) and the results from the model fits (effective diameter $D_{eff}$, geometric albedo $p_v$ and beaming factor $\eta$), are shown in **Figure 7**. The STM and NEATM models both fit the IRAS measurements; the NEATM model fit is significantly better, however, and it provides size estimates closer to our size model, $D_{eff}$= 154.1 km & 153.6 km, with a conservative error of 15 km. Using our pole solution, the non-convex model and our equivalent size ($D_{eq} = 154 \pm 6$ km) for the primary, we calculated the projected equivalent diameter for each IRAS sighting. The sub-Earth latitude of Minerva at the time of the two IRAS sightings was +23 deg, and the corresponding predicted effective diameters (155.5 km & 158.3 km) were in very good agreement with the NEATM results.
From our equivalent diameter and considering the geometric albedo $p_v = 0.062 \pm 0.013$ calculated with NEATM, we derive an absolute visible magnitude for the asteroid $H_v = 7.7 \pm 0.2$ in agreement with the currently reported value in the JPL HORIZONS ephemeris.
**[INSERT HERE FIG. 7]**

## 4 Mutual Orbits of the moonlets
### 4.1 Determination of the mutual Orbit

In Table 2 we report the measured relative positions of satellites with respect to the center-of-light of Minerva from the AO observations collected in August and September 2009. These astrometric positions were derived after fitting the moonlet and primary centroid with a Moffat-Gauss 2D stellar profile from a specific method suited to adaptive optics images (Descamps et al., 2011). The 1-sigma error on the position of the satellites is estimated to be 5-7 *mas*. Because we also derived an accurate shape of Minerva (Section 3), we can also estimate the error due to the phase angle, which is less than 3 *mas*. The total astrometric error on the position of the satellites is 8-10 *mas*.

We estimated the mutual orbits of the satellites by fitting them independently using our genetic-based algorithm, Genoide-Kepler, extensively described and validated in Vachier et al. (2012). This computationally efficient algorithm provides a Keplerian orbital solution for the assumed mass-less satellite of a primary asteroid, which is treated as a point mass. Based on a statistical method, this algorithm searches for the eight parameters describing a mutual orbit without a priori information. We initiated a search of each mutual orbit by running approximately $1 \times 10^6$ trials and refining the analysis around the best solution. The quality of the solution is defined by its fitness function, $f_p$, defined as the quadratic mean of observed minus computer positions. Orbital solutions have been found for Minerva I ($f_p \sim 10$ *mas*) and Minerva II ($f_p \sim 3$ *mas*). The Minerva II solution has a better fitness function and is certainly more accurate since it is based on 10 astrometric positions. **Table 4** provides the osculating elements of the satellite orbits. The 3-sigma errors on each element is estimated by considering an uncertainty on the astrometric positions of 10 *mas* and searching for the solution which provides a fitness function $f_p < 31$ *mas*. Supplement materials S1 & S2 contain information on the numerical analysis and the estimate of the orbital parameters for each satellite.

Both orbits are coplanar, having a difference in inclination of 7 deg, but not aligned with the equator pole of the primary (with $\Delta i = |i_{sat} - i_{equator}| \sim 42$ deg for Minerva II). However, the 3-sigma error is quite high (~33 deg) implying that the system inclination is poorly constrained due to the almost face-on geometry of the system in the 2010 opposition. The highest eccentricity was measured for Minerva II (e=0.05 with a 1-sigma error of 0.04), so we can conclude that no orbital eccentricity is measurable.

 [INSERT HERE TABLE 4]

### 4.2 Comparison with other main-belt triple asteroid systems

A comparison of the architectures of all known main-belt triple asteroid systems is shown in **Fig. 8**. The characteristics of these multiple systems come from several sources. The orbital parameters of the (45) Eugenia satellites were determined using a dynamical model presented in Marchis et al. (2010). (87) Sylvia, the first known triple asteroid, had its mutual orbits determined in Marchis et al. (2005) using a geometrical model and in Fang et (2012) using a more complex dynamical model and a longer baseline of observations. Both models are in agreement for the major axis, period and eccentricity but differ for the estimate of the $J_2$. The mutual orbits of the (216) Kleopatra satellites were approximated using a geometrical model in Descamps et al. (2011). The case of the triple asteroid (3749) Balam system is more complex, since this triple system is not yet fully constrained. Vachier et al. (2012) showed the existence of 10 orbital solutions with equal fitness quality for its outer satellite. We chose one of these solutions ("sol 4" in Table 2 of Vachier et al. (2012) for the schematic in **Fig. 8** and the new size estimate of the components based on Spitzer/IRS observations (Marchis et. al 2012a) combined with Polishook et al. (2011) photometric survey. In the case of Balam's inner satellite, we used the orbital elements provided in Marchis et al. (2008c). **Table 5** lists the parameters used as inputs in the creation of Figure 8. They were extracted from the references mentioned above and the JPL HORIZONS ephemeris (to calculate the radius of the Hill sphere).
**[INSERT HERE TABLE 5]**

The large triple systems (45) Eugenia, (87) Sylvia and (216) Kleopatra are similar in many ways to (93) Minerva. They are composed of a large ($D_p > 130$ km) primary surrounded with small satellites ($D_s = 3 - 18$ km). The satellites orbit well inside the Hill sphere ($a = 1\text{-}3\% \times R_{Hill}$) and close to the primary ($a = 5 - 12 \times R_p$). For all satellites except Eugenia-II ($e \sim 0.069$), their orbits are almost circular ($e < 0.007$) suggesting damping by tidal effects.

For these large triple asteroid systems, the satellites' orbits have the same direction as the spin of the primary. In the case of (87) Sylvia, (93) Minerva and (216) Kleopatra, the satellites' orbits are also coplanar. The satellites of (45) Eugenia are significantly inclined with respect to each other ($\Delta i_{sat} = |i_{sat-I} - i_{sat-II}| \sim 9$ deg) and with respect to the equator of the primary ($\Delta i = | i_{sat} - i_{equator} | \sim 18$ and 9 deg for the inner and outer satellites respectively). By studying the long-term evolution of the orbits, Marchis et al. (2010) showed that these orbits are affected by solar perturbations and, to a lesser extent, by mutual interaction between the two moons. The case of (93) Minerva is complex. In Section 4.1 we showed that even if the orbits of the satellite are coplanar ($\Delta i_{sat} \sim 7$ deg), they are significantly inclined with respect to the equator of the primary ($\Delta i = 42$ deg for Minerva II). However, since the system was pole-on at the time of the observations, the satellite inclination is poorly constrained with a 3-sigma error of approximately 35 deg. Additional observations taken during future oppositions and with different geometries, will refine the inclination of the satellites' orbits with respect to the primary equator.

## 5 Composition of the System
### 5.1 Bulk density of (93) Minerva
After combining the two masses obtained individually from the orbits each of Minerva's moons, we found a total mass of the system M = $3.35 \pm 0.54 \times 10^{18}$ kg. This gives an average density of $\rho = 1.75 \pm 0.30$ g/cm$^3$ adding the 1-sigma error on the mass quadratically and propagating the 1-sigma error on the size. This density is higher than the density of multiple C-complex asteroids with a similar primary size reported in Marchis et al. (2008a, 2008b) such as (45) Eugenia ($\rho = 1.1 \pm 0.1$ g/cm$^3$, $D_p \sim 215$ km), (379) Huenna ($\rho = 0.85 \pm 0.05$ g/cm$^3$, $D_p \sim 215$ km), (762) Pulcova ($\rho = 0.9 \pm 0.1$ g/cm$^3$, $D_p \sim 140$ km) and in the upper limit of (121) Hermione ($\rho = 1.4$ +0.5/-0.2 g/cm$^3$, $D_p \sim 187$ km) in Descamps et al. (2009). The average density of (93) Minerva is slightly higher than those other C-type asteroids suggesting that it could have a different composition.

### 5.2 Reflectance Spectroscopy
We observed (93) Minerva on September 13, 2010 at 15:19 UT using the SpeX instrument available on the NASA 3m InfraRed Telescope Facility (IRTF) atop Mauna Kea, Hawaii. With an estimated brightness of V~13.7, the multiple asteroid was bright enough to collect the high signal-to-noise ratio (S/N) reflectance spectrum necessary to refine the taxonomic type of (93) Minerva. This spectrum was obtained following the procedures described in Emery et al. (2005). Data frames were taken in pairs with the object dithered along the slit. Subtraction of these pairs produces a first-order removal of sky emission. The on-chip integration time was 120s to minimize the effects of variability of the sky emission. For each frame, we applied a linearity correction. Typically, multiple frames were taken with the maximum on-chip integration time and summed together during data reduction in order to increase the S/N ratio. Several nearby solar analog stars were observed regularly throughout the asteroid observations to ensure good correction of atmospheric absorption and solar spectral slope. We checked the quality of Minerva's solar analog star (HD260033) by cross-comparison with observations of other solar analogs recorded the same night. With SpeX, flat field images are obtained by illuminating an integrating sphere, which is in a calibration box attached to the spectrograph. This box also contains an argon lamp that is used for wavelength calibration. The resulting reflectance spectrum of the (93) Minerva system is shown in **Figure 9.**

(93) Minerva was also observed on November 1, 2009 at 00:47 UT from the 3.6m Telescopio Nazionale Galileo (TNG) located on the Island of San Miguel de La Palma using the Near Infrared Camera Spectrometer (NICS). Ambient conditions were poor, with a mostly clear sky but highly variable seeing ranging from 0.8" to 2.0". The spectrometer was used with a 1.5" slit and a JK' grism covering the wavelength range between 1.2 and 2.2 µm, with a central resolution of ~300. Observations were made following a standard ABBA sequence to accurately subtract the sky. Two successive ABBA sequences were obtained with a total integration time of 30s at each position sufficient to provide a good S/N ratio for the V = 11.8 target. The solar analog HIP4113 (G0) was observed just after Minerva using the same settings but an integration time of only 3s at each position. The data were processed using standard procedures described above, including flat field correction, distortion corrections, sky-subtraction using consecutive pairs of images, co-addition of the individual images and extraction of the spectrum. A wavelength solution was computed using the spectrum of Ar-Xe lamps. The TNG spectrum displayed in Fig. 9 has a lesser quality than the IRTF spectrum due to the poorer atmospheric conditions from Roque de los Muchachos Observatory (alt. ~ 2,396 m) in comparison with Mauna Kea (alt. ~ 4,207 m). Using the shape model derived in Section 3, we calculate the appearance of the primary (Fig. 9) and show that two different areas of the primary asteroid have been observed. The slopes of these two spectra are remarkably identical, suggesting that (93) Minerva has a homogenous surface composition.

To the recently collected NIR spectra, we added a visible spectrum from SMASSII database (Bus & Binzel, 2002a). Based on the visible reflectance spectrum, Bus & Binzel (2002b) sorted (93) Minerva as a C-type asteroid (weak/medium UV band shortward of 0.55 µm, flat or slightly red and featureless longward of 0.55 µm) in agreement with its low-albedo derived from our reanalysis of IRAS observations ($p_v$ ~ 0.062 ± 0.013, see Figure 7). The combined visible and near-infrared spectra of (93) Minerva shown in Fig. 9 were analyzed using the method described in deMeo et al. (2009). With the exception of a slightly redder slope seen only on the IRTF spectrum, (93) Minerva spectra are similar to the average C-type asteroids of the deMeo et al. (2009) taxonomy. It is worth noting that the comparison of asteroids based solely on their visible and near-infrared spectra could be flawed. For example, the presence of a UV absorption band at 0.4 μm differentiated the P- (no absorption) and G- (strong absorption) class in the Tholen & Barucci (1989) taxonomy. It is therefore possible that the C-class is in fact composed of several asteroids with identical visible and near-infrared spectra but a different composition.

The relatively high density $\rho = 1.75 \pm 0.30$ g/cm$^3$ of (93) Minerva is similar to the density of several G-type asteroids of the Tholen & Barucci (1989) taxonomy like (130) Elektra ($\rho = 2.1/1.7 \pm 0.3$ g/cm$^3$ in Marchis et al. 2008a). The bulk density of (1) Ceres, also classified as member of the G-type family, is estimated to $2.0 \pm 0.3$ g/cm$^3$ in Britt et al. (2002). However we have to be cautious in this case Ceres' interior should be able to re-compact itself by its own gravity due to its large size (D~945 km). Therefore its bulk density is not comparable with the one estimated for ~100 km size asteroids. Unfortunately, these are the only density estimates of G-type asteroids available so far. Additional observations of (93) Minerva, including spectroscopy in the UV and mid-IR observations, could shed light on the difference in composition, hence density, of C-complex asteroids.

### 5.3 Meteorite analogues and macroporosity

By comparing Minerva's average density to those of meteorites, it is possible to constrain the macroporosity and internal composition of (93) Minerva. Using the density and porosity data listed in Britt et al. (2006) and assuming that Minerva has a pure composition of one type of meteorite, we derived the macroporosity of the asteroid.

CI and CM carbonaceous chondrite meteorites which show high components of hydrated material typically have densities of approximately 2.11 g/cm$^3$ and average porosities between 11 and 23%. Consequently, if Minerva was made of the same material, its macroporosity will be less than 10%, implying that the asteroid is coherent and slightly fractured. For the dry CR, CO, and CV meteorites with a density of ~3 g/cm$^3$ and a porosity of 14 to 20%, the macroporosity is estimated to be 30%. Similar porosities were derived for several binary asteroids (e.g. Marchis et al. 2008a, Marchis et al. 2008b) and for single asteroids visited by spacecraft (Mathilde, Lutetia and Itokawa, see a review in Barucci et al. 2011). The shape of craters on Mathilde (Veverka et al. 1997) and the absence of impact craters on Itokawa (Fukiwara et al. 2006) observed on close-up images collected by spacecraft implied that these asteroids have a rubble-pile interior. We conclude that if (93) Minerva is similar to all observed multiple asteroids, then Minerva's primary is a rubble pile asteroid; it might be made of dry carbonaceous chondrite meteorite material.

### 6. Origin of the system

In Section 3, we derived an accurate size ($D_{eq} = 154 \pm 6$ km) and spherical shape model of Minerva's primary. In Section 4 we showed that the moons' mutual size ratio and orbits have more similar characteristics than those derived for large multiple asteroids in the main-belt. In Section 5, we found that this C-type asteroid has a high density of $1.75 \pm 0.30$ g/cm$^3$, similar to all G-type asteroids, implying that the C-class of deMeo et al. (2009) is likely made of asteroids with different compositions. (93) Minerva could be a rubble-pile asteroid, like other multiple asteroids, if it has the same composition as dry carbonaceous chondrites.

Without having a clear idea of the composition of this asteroid and due to the lack of high-resolution images, it is difficult to assess the origin of the system. However, since both (93) Minerva and (45) Eugenia (Marchis et al. 2010) display similarities, we can envision that (93)

Minerva also formed as the result of a catastrophic disruption of a parent asteroid (Merline et al. 1999, Marchis et al. 2010) or an ejection of fragments after an oblique impact (Descamps & Marchis, 2008).

Considering the tidal dissipation between the satellites and the primary and assuming that the satellite was formed outside the synchronous orbit, we can estimate the tidal evolution timescale of the satellites. The principles of tidal evolution for binary asteroid system are discussed in Marchis et al. (2008a). The tidal parameter μQ, the product of rigidity (μ) and specific dissipation (Q), is the key parameter to estimate the age of the system. Assuming $\mu = 10^9$ N/m$^2$ and Q~100, derived for the C-type binary asteroid (283) Emma in Marchis et al. (2008a) and using the orbital elements of the outer satellite, the age of the (93) Minerva system is estimated to ~1 billion of years. However, it is possible that the age of the system is underestimated since the system could have been or could still be in an excited orbit, implying that the orbit has not been evolving due to tidal effects only. Consequently, the (93) Minerva system may be significantly older. Vokrouhlický et al. (2010) discussed the apparent paradox of binary asteroids being remnants of disruptive collisions and the absence of detectable collisional families for Cybele binary asteroids. Their work suggests that Cybele binary systems, (121) Hermione and (107) Camilla, formed ~4 billion years ago and that the effacement of these families is linked to resonance sweeping associated with Jupiter's late inward migration. A similar study for multiple asteroids formed by catastrophic disruption and located in the middle of the main-belt, such as (93) Minerva, is needed to assess if most of the large (D > 100 km) high-size ratio binary asteroids (called T1 in Marchis et al. (2012a) nomenclature) are indeed witnesses of effaced families.

Several different scenarios of formation have been proposed since the discovery of T1 asteroids. Descamps and Marchis (2008) postulated that since all T1 binaries are close to the stability limit of self-gravitating, rotating ellipsoids made of cohesionless granular materials, a mass shedding process could have preceded the formation of the satellite. Based on Descamps and Marchis (2008) and from the non-convex primary shape (with a fitted ellipsoid of axes *a₀ = 1.15 × a₁ = 1.25 × a₂* and a dimensional parameter which quantifies the non-sphericity of the primary $\lambda_p$ ~ 1.46) and assumed spherical satellites ($\lambda_s$ ~ 1), we calculated the specific angular momentum $H_m$ = 0.23 ± 0.02 for both satellites and a normalized spin rate $\Omega$ = 0.40 ± 0.02. (93) Minerva marginally clusters with other T1 binaries shown in Fig. 1 of Descamps and Marchis (2008). Its scaled spin rate and specific angular momentum are too low to place it at the Maclaurin-Jacobi transition like other T1 asteroids described in this work. This implies that it could have formed differently than other T1 asteroids or that the orbits of its satellites may not be primordial and are instead in a not-yet identified excited state.

## 7. Summary of findings and future work
(93) Minerva is a multiple asteroid in the family of known high-sized ratio large asteroid systems (T1). It shares part of the characteristics of this population of multiple asteroid systems, such as a large (D = 154 ± 6 km) primary surrounded by two small (Deq~ 3 km) moons orbiting very close to the primary and well-inside the Hill sphere ($a_I$ = 8.0 × $R_p$ = 2.1% × $R_{Hill}$ & $a_{II}$ = 4.8 × $R_p$ = 1.2% × $R_{Hill}$) in an almost-circular orbit. It has, however, some unique characteristics, which make it an interesting and puzzling body in this sub-population. Specifically:

- An almost circular shape for the primary approximated by an ellipsoid with axis $a_0 = 1.15 \times a_1 = 1.25 \times a_2$, leading to a quadrupole moment $J_2 \sim 0.08$ assuming a homogenous distribution of mass in its interior.
- A possible inclination of the satellite orbits up to 42 deg, not seen yet for any binary asteroids. However, this could be due to an almost face-on measurements and the lack of astrometric positions since 3-sigma error on the inclination is ~33 deg. If this inclination is confirmed to be real, it suggests that the orbits of the satellites are in excited states.
- A high density of $1.75 \pm 0.30$ g/cm$^3$, suggesting that the C-class in deMeo et al. (2009) taxonomy is composed of asteroids with different compositions, since several binary C-type asteroids have lower densities than (93) Minerva. The G-type asteroid (130) Elektra share a similar high density, so Minerva may be similar in composition to dry CR, CO, and CV meteorites.
- (93) Minerva is not located at the Maclaurin-Jacobi transition of a cohesionless rotating fluid like other T1 asteroids. Hence its multiplicity may not be due to mass shedding after an oblique impact or any process spinning up the parent primary asteroid. A catastrophic disruption of a parent asteroid, another proposed scenario, is possible if the event occurred several billion years ago and if a process is identified to efface all clues about an early "Minerva" collisional family.

This work provides accurate insights into the properties of (93) Minerva, the fourth triple asteroid in the main-belt to have both moons imaged. The characteristics of the mutual orbits and shape and size of the primary are well constrained thanks to tools we developed over the past several years to analyze single and binary asteroids.

Our understanding of the detailed nature of (93) Minerva, such as its composition and internal structure, as well as the formation and evolution of its satellites is still incomplete. Future adaptive optics observations at exceptionally close oppositions (distance to Earth ~ 1.4 AU) such as in August 2013 (V=10.8) and in May 2017 (V=10.9) will help refine the orbits of the satellites, confirming their significant inclination and detecting possible resonances between the components of the system. Observation campaigns with a large number of stations (e.g. (90) Antiope by Colas et al. 2011) to detect stellar occultations by the primary asteroid and its satellites are a relatively low-cost way to constrain the size and shape of the primary and the satellites without ambiguity.

In this work we discussed the limit of the visible and near-infrared spectra to characterize the surface of C-type asteroids. Expanding the spectroscopic study to mid-infrared and eventually UV will be useful in characterizing the surface properties of (93) Minerva through comparison with meteorite lab sample spectra. The detection of emission features in the mid-infrared spectra of numerous large and binary asteroids by Marchis et al., (2012a) constrained the nature of asteroid surfaces (e.g. Vernazza et al., 2012). Future facilities and instruments, providing access to a broad wavelength range and exceptionally high angular resolution, linked with laboratory measurements is essential to the understanding of multiple asteroids, witnesses of the collisional past and the complex history of our solar system.

**Acknowledgements:**
FMA and JEM were supported by NASA grant NNX11AD62G. Parts of these data were obtained with the W.M. Keck Observatory, which is operated by the California Institute of Technology, the University of California, Berkeley and the National Aeronautics and Space Administration. The work of JD was supported by grants P209/10/0537 of the Czech Science Foundation and by the Research Program MSM0021620860 of the Ministry of Education. We are thankful to H.-Y. Shih and K. Larson for their participation in the photometric observations. Thanks to D. Polishook and an anonymous reviewer for their comments which improved significantly the quality of this work. The observatory was made possible by the generous financial support of the W.M. Keck Foundation. The authors extend special thanks to those of Hawaiian ancestry on whose sacred mountain we are privileged to be guests. Without their generous hospitality, none of the observations presented would have been possible. The authors would like to thank the IOTA group for helping to coordinate and gather observations of the December 24 2010 stellar occultation especially the observers: R. Peterson, G. Lucas, J. Ray, S. Herchak, J. Menke, W. Thomas, D.J Dunham , B. Jones, S. Conard.

**Table 1:** Conditions of the Keck II AO observations of (93) Minerva collected in 2009.

| Date UT | Time UT | Filter | airmass | predicted V | Elongation (deg.) | α Phase (deg) | Δ Distance to Earth (AU) |
|---|---|---|---|---|---|---|---|
| 16-Aug-09 | 13:38:24 | Fe II | 1.03 | 12.47 | 114.69 | 19.95 | 2.110850 |
| 16-Aug-09 | 13:45:02 | Kc | 1.03 | 12.47 | 114.70 | 19.96 | 2.110799 |
| 16-Aug-09 | 13:56:12 | Fe II | 1.02 | 12.47 | 114.71 | 19.96 | 2.110719 |
| 16-Aug-09 | 14:25:02 | Fe II | 1.01 | 12.47 | 114.72 | 19.95 | 2.110509 |
| 16-Aug-09 | 14:40:54 | Fe II | 1.01 | 12.47 | 114.73 | 19.95 | 2.110401 |
| 16-Aug-09 | 15:29:45 | Fe II | 1.04 | 12.47 | 114.77 | 19.94 | 2.110048 |
| 06-Sep-09 | 15:30:58 | Fe II | 1.23 | 12.13 | 135.32 | 15.07 | 1.916779 |
| 12-Sep-09 | 15:10:47 | Fe II | 1.26 | 12.02 | 141.72 | 13.17 | 1.874003 |
| 15-Sep-09 | 15:11:48 | Fe II | 1.32 | 11.96 | 145.02 | 12.15 | 1.855197 |
| 28-Sep-09 | 13:34:08 | Fe II | 1.18 | 11.71 | 159.74 | 7.22 | 1.797652 |

**Table 2:** Astrometric positions of Minerva's moonlets extracted from W.M Keck AO observations taken in 2009 with the NIRC-2 camera. The diameters of the moons are estimated by assuming the same albedo, and therefore surface composition, between the primary and the satellites.

| Date UT | Time UT | JD -2455000 | satellite | Filter | X arcsec | Y arcsec | Diameter km |
|---|---|---|---|---|---|---|---|
| 16-Aug-09 | 13:38:24 | 60.0683333 | Minerva I | Fe II | -0.404 | 0.034 | 4.1 |
| 16-Aug-09 | 13:45:02 | 60.0729398 | Minerva I | Kc | -0.401 | 0.020 | 4.0 |
| 16-Aug-09 | 13:56:12 | 60.0806944 | Minerva I | Fe II | -0.407 | 0.015 | 3.6 |
| 16-Aug-09 | 14:24:01 | 60.1007175 | Minerva I | Fe II | -0.400 | -0.002 | 2.6 |
| 16-Aug-09 | 14:40:54 | 60.1117361 | Minerva I | Fe II | -0.397 | -0.017 | 2.7 |
| 16-Aug-09 | 15:30:58 | 60.1456590 | Minerva I | Fe II | -0.407 | -0.059 | 2.5 |
| 6-Sep-09 | 15:30:58 | 81.1465082 | Minerva I | Fe II | 0.005 | 0.441 | 2.7 |
| 12-Sep-09 | 15:10:47 | 87.1324920 | Minerva I | Fe II | -0.069 | -0.435 | 3.7 |
| 15-Sep-09 | 15:11:48 | 90.1331924 | Minerva I | Fe II | 0.428 | -0.027 | 4.9 |
| 28-Sep-09 | 13:34:08 | 103.0653659 | Minerva I | Fe II | -0.308 | 0.365 | 5.1 |
| 16-Aug-09 | 13:56:12 | 60.0806944 | Minerva II | Fe II | -0.148 | -0.204 | 4.1 |
| 16-Aug-09 | 14:40:54 | 60.1117360 | Minerva II | Fe II | -0.118 | -0.216 | 3.2 |
| 6-Sep-09 | 15:30:58 | 81.1465082 | Minerva II | Fe II | -0.261 | -0.094 | 1.6 |
| 12-Sep-09 | 15:10:47 | 87.1324920 | Minerva II | Fe II | 0.220 | -0.102 | 3.2 |
| 15-Sep-09 | 15:11:48 | 90.1331924 | Minerva II | Fe II | -0.214 | -0.191 | 4.0 |
| 28-Sep-09 | 13:34:08 | 103.0653659 | Minerva II | Fe II | 0.026 | 0.280 | 3.2 |

**Table 3:** Observation circumstances of the photometric lightcurve observations of (93) Minerva used in the shape inversion process described in Section 4.

| date | r | delta | phase | observer |
|---|---|---|---|---|
| | AU | AU | deg | or reference |
| 1980 04 11.3 | 2.717 | 1.768 | 8.5 | Harris and Young 1989 |
| 1981 09 16.2 | 2.469 | 1.642 | 16.3 | Debehogne et al. 1982 |
| 1981 09 19.1 | 2.472 | 1.672 | 17.2 | Debehogne et al. 1982 |
| 1982 10 18.3 | 2.993 | 2.173 | 12.7 | Harris and Young 1999 |
| 1982 12 14.2 | 3.047 | 2.152 | 9.2 | Harris and Young 1999 |
| 1982 12 14.3 | 3.047 | 2.153 | 9.2 | Harris and Young 1999 |
| 1982 12 15.2 | 3.048 | 2.159 | 9.5 | Harris and Young 1999 |
| 1982 12 19.2 | 3.051 | 2.193 | 10.7 | Harris and Young 1999 |
| 1983 01 08.2 | 3.067 | 2.410 | 15.5 | Harris and Young 1999 |
| 1983 01 16.2 | 3.073 | 2.514 | 16.7 | Harris and Young 1999 |
| 1998 02 21.9 | 2.924 | 1.940 | 2.4 | Denchev 2000 |
| 2005 12 22.8 | 3.091 | 2.188 | 8.6 | Torppa et al. 2008 |
| 2006 01 14.7 | 3.105 | 2.406 | 14.5 | Torppa et al. 2008 |
| 2006 12 25.0 | 3.054 | 2.421 | 15.9 | Torppa et al. 2008 |
| 2007 01 30.9 | 3.021 | 2.069 | 5.9 | Torppa et al. 2008 |
| 2007 02 21.8 | 2.998 | 2.028 | 4.4 | Torppa et al. 2008 |
| 2009 11 08.1 | 2.818 | 1.911 | 9.9 | Alton 2011 |
| 2009 11 09.1 | 2.820 | 1.920 | 10.2 | Alton 2011 |
| 2009 11 18.0 | 2.832 | 2.005 | 13.0 | Alton 2011 |
| 2009 11 22.1 | 2.838 | 2.049 | 14.2 | Alton 2011 |
| 2009 11 25.1 | 2.842 | 2.084 | 14.9 | Alton 2011 |
| 2009 12 02.1 | 2.852 | 2.171 | 16.5 | Alton 2011 |
| 2011 02 04.0 | 3.140 | 2.331 | 11.9 | Strajnic |
| 2011 03 05.3 | 3.133 | 2.661 | 17.4 | Marchis |
| 2011 03 06.2 | 3.133 | 2.673 | 17.5 | Marchis |
| 2011 03 07.2 | 3.132 | 2.685 | 17.6 | Marchis |
| 2011 03 09.2 | 3.132 | 2.712 | 17.8 | Marchis |
| 2011 03 10.2 | 3.131 | 2.726 | 17.8 | Marchis |
| 2011 03 11.2 | 3.131 | 2.739 | 17.9 | Marchis |
| 2012 01 11.4 | 2.854 | 2.452 | 19.5 | Dalba |
| 2012 02 18.4 | 2.801 | 1.956 | 12.6 | Dalba |

**Table 4:** Minerva system characteristics and osculating elements of Minerva I (outer satellite) and Minerva II (inner satellite) orbits.

| Component | Parameter | Value | 1-$\sigma$ Error | Units |
|---|---|---|---|---|
| Minerva | Mass | 3.35 | 0.54 | $10^{18}$ kg |
| | Equivalent Diameter | 154 | 6 | km |
| | Absolute visible magnitude | 7.7 | 0.2 | |
| | Geometrical albedo | 0.062 | 0.013 | |
| | Bulk density | 1.75 | 0.30 | g/cm$^3$ |
| | $J_2$ | 0.08 | | |
| | Rotation period | 5.981767 | 0.000004 | hours |
| | Pole solution in ecliptic ECJ2000 | $\lambda$=21 | 10 | degrees |
| | | $\beta$=21 | 10 | |
| Minerva I | Mass | Negligible | Set to 0 | |
| | Estimated diameter | 3.6 | 1.0 | km |
| | Semimajor axis | 623.5 | 10 | km |
| | Period | 2.406 | 0.002 | days |
| | Eccentricity | 0 | 0.009 | |
| | Inclination in EQJ2000 | 89.0 | 10.9 | degrees |
| | Longitude of the Ascending node in EQJ2000 | 126.0 | 10.0 | degrees |
| | Argument of periapsis in EQJ2000 | 82.0 | 14.4 | degrees |
| | Pericenter date | 2455059.4 | 0.10 | days |
| | Pole orientation in ecliptic ECJ2000 | $\lambda$=34 | 10 | degrees |
| | | $\beta$=-13 | 10 | |
| Minerva II | Mass | Negligible | Set to 0 | |
| | Estimated diameter | 3.2 | 0.9 | km |
| | Semimajor axis | 375 | 16 | km |
| | Period | 1.1147 | 0.0006 | days |
| | Eccentricity | 0.05 | 0.04 | |
| | Inclination in EQJ2000 | 91.4 | 24/1 | degrees |
| | Longitude of the Ascending node in EQJ2000 | 132.6 | 23.6 | degrees |
| | Argument of periapsis in EQJ2000 | 347.5 | 27.7 | degrees |
| | Pericenter date | 2455059.30 | 0.08 | days |
| | Pole orientation in ecliptic ECJ2000 | $\lambda$=40 | 20 | degrees |
| | | $\beta$=-17 | 28 | |

**Table 5:** Orbital parameters (*a* semi-major axis, *P* period, *e* eccentricity) and physical characteristics ($R_p$ radius of the primary, $R_s$ radius of the satellite, $R_{Hill}$ radius of the Hill sphere) of the known triple asteroid systems located in the main-belt. The size of Balam components was derived from Spitzer/IRS observations in Marchis et al. (2012a).

| ID | Name | a | P | e | $R_p$ | $R_s$ | err $R_s$ | $R_{Hill}$ | $a/R_p$ | $a/R_{Hill}$ | reference |
|---|---|---|---|---|---|---|---|---|---|---|---|
| | | km | days | | km | km | km | km | | | |
| 45 | Eugenia-I | 1164 | 4.703 | 0.006 | 95 | 7 | 1 | 40400 | 12.3 | 0.0288 | Marchis et al. 2010 |
| 45 | Eugenia-II | 611 | 1.787 | 0.069 | 95 | 5 | 1 | 40400 | 6.4 | 0.0151 | Marchis et al. 2010 |
| 87 | Sylvia-I | 1356 | 3.6496 | 0.000 | 143 | 18 | 2 | 70700 | 9.5 | 0.0192 | Marchis et al. 2005 |
| 87 | Sylvia-II | 706 | 1.3788 | 0.000 | 143 | 7 | 1 | 70700 | 4.9 | 0.0100 | Marchis et al. 2005 |
| 93 | Minerva-I | 623 | 2.406 | 0.001 | 78 | 4 | 1 | 30416 | 8.0 | 0.0205 | this work |
| 93 | Minerva-II | 375 | 1.1147 | 0.050 | 78 | 3 | 1 | 30416 | 4.8 | 0.0123 | this work |
| 216 | Kleopatra-I | 678 | 2.32 | 0.001 | 67.5 | 9 | 2 | 38454 | 10.0 | 0.0176 | Descamps et al. 2011 |
| 216 | Kleopatra-II | 454 | 1.24 | 0.001 | 67.5 | 7 | 2 | 38454 | 6.7 | 0.0118 | Descamps et al. 2011 |
| 3749 | Balam-I | 203.4 | 81.43 | 0.573 | 2.15 | 0.9 | 0.3 | 859 | 94.6 | 0.2368 | Vachier et al. 2012 |
| 3749 | Balam-II | 20 | 1.391 | 0.000 | 2.15 | 0.85 | 0.3 | 859 | 9.3 | 0.0233 | Marchis et al. 2008 |

**Figure 1:** Observations of (93) Minerva recorded on August 16, 2009 with the W.M. Keck-II telescope and its adaptive optics. Five of these observations were recorded with the FeII filter centered at $\lambda_c$ = 1.6455 µm. The processed frame taken at 13:45 UT was recorded using the Kcont filter centered at $\lambda_c$ = 2.2706 µm. The horizontal arrow indicated the position of the outer moonlet "Minerva I" seen on the frame recorded at 13:38 UT. The vertical arrow corresponds to the position of the inner moonlet "Minerva II" detected on the frame taken at 13:56 UT. The position of each arrow is fixed on the images, hence it shows the motion of the moonlets over ~2 and ~1h of observation for Minerva I and Minerva II respectively.

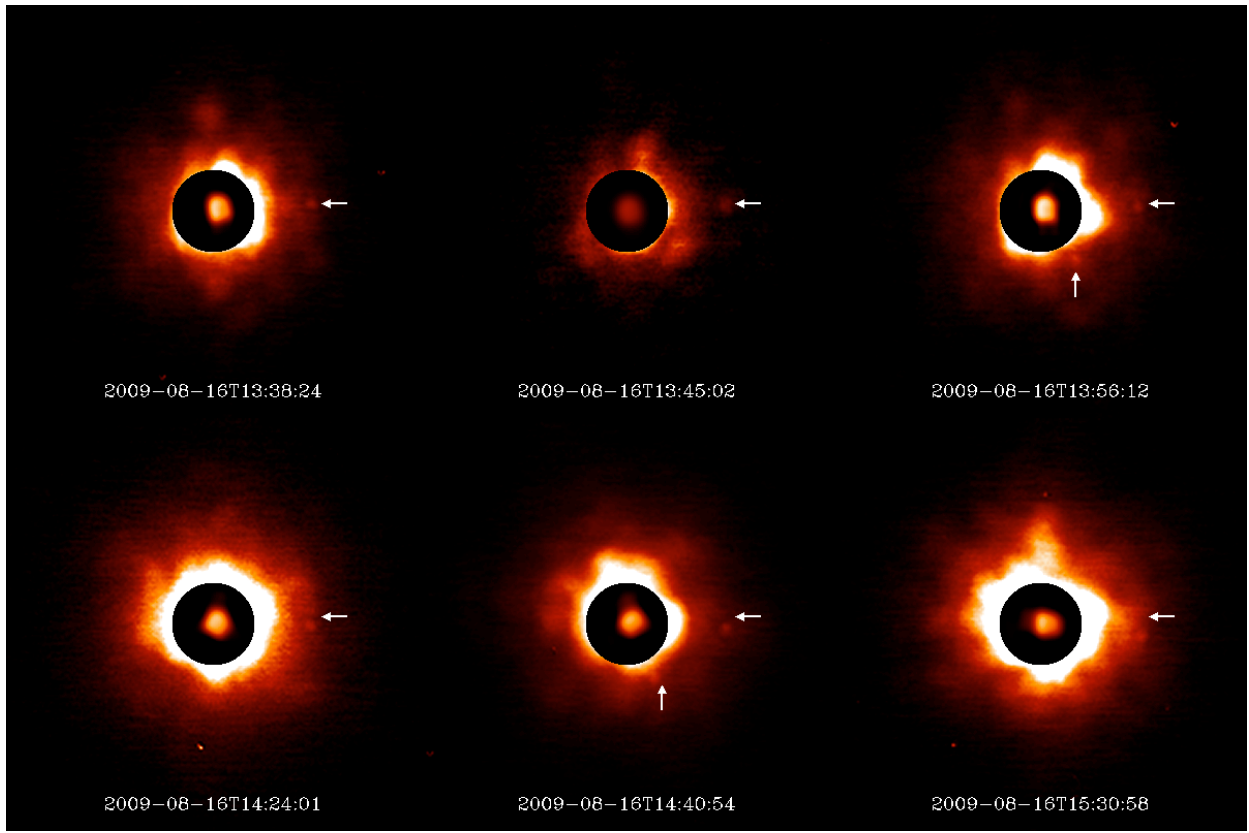

**Figure 2:** Additional W.M. Keck II AO observations of (93) Minerva and its two moonlets were recorded in September 2009 using an FeII band filter. The positions of the moonlets are indicated with horizontal and vertical arrows for Minerva I and Minerva II respectively. The variable quality of the data provided by the AO system is visible. For instance, on September 6 the two satellites are barely visible since their brightness is very close to the uncorrected halo surrounding the primary. The intensity of this halo is reduced for the image taken on September 15, improving the contrast on the satellite signals.

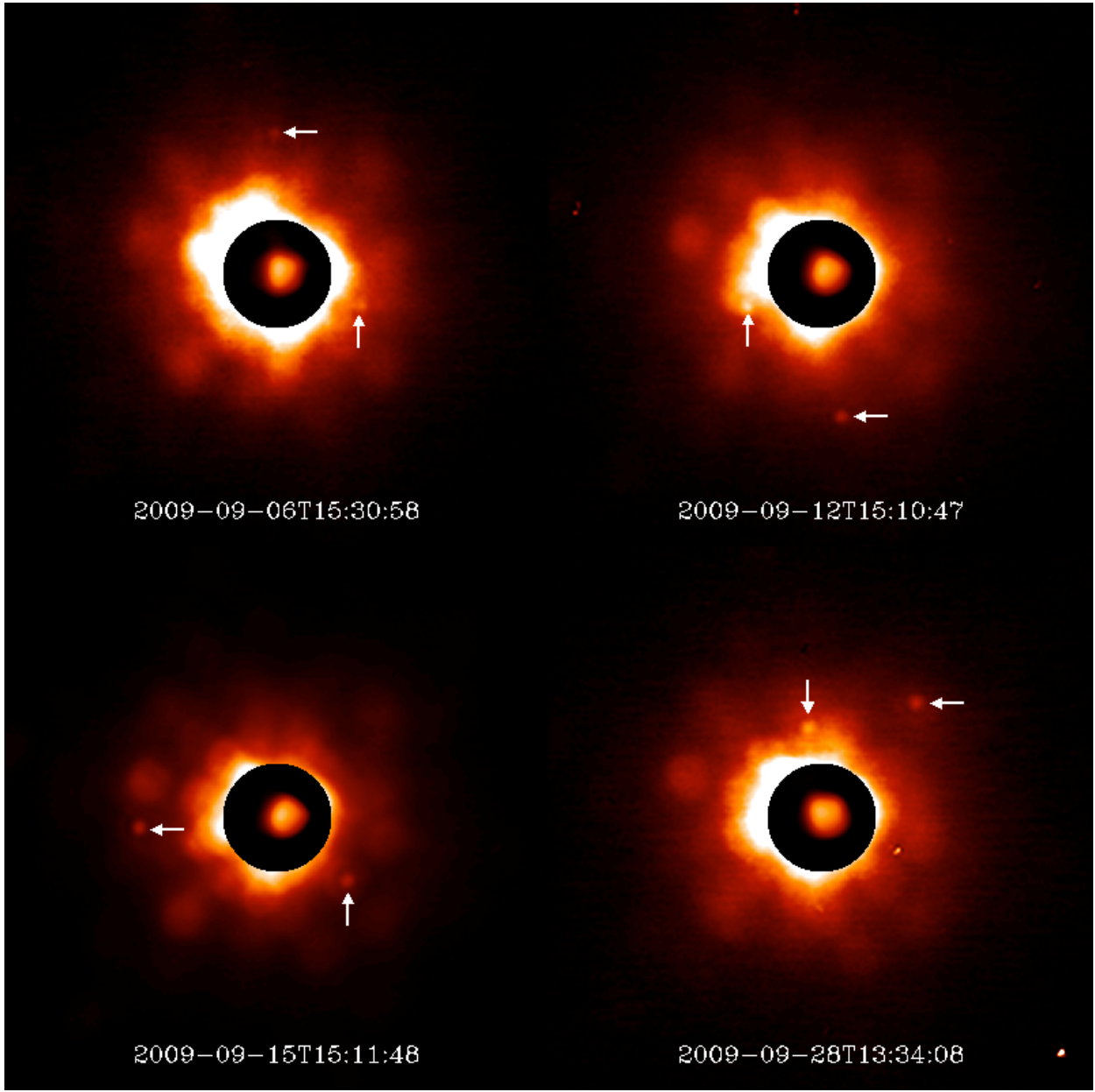

**Figure 3:** Observed (dots) photometric lightcurves of (93) Minerva collected from 1980 to 2012 in the optical wavelength range and displayed in relative intensity. The solid line represents the fit of our 3D shape model. θ and θ$_o$ are the aspect angles of Earth and the Sun respectively. α is the phase angle. These lightcurves have small amplitudes of less than 0.25 mag, even when the asteroid was observed near the equator of the primary (θ ~ θ$_o$ ~90 deg) between February and March 2011, implying that the asteroid is regular in shape. The last frame, taken on Feb 18 2012, shows some discrepancies due to photometric error introduced by the poor imaging quality during the night.

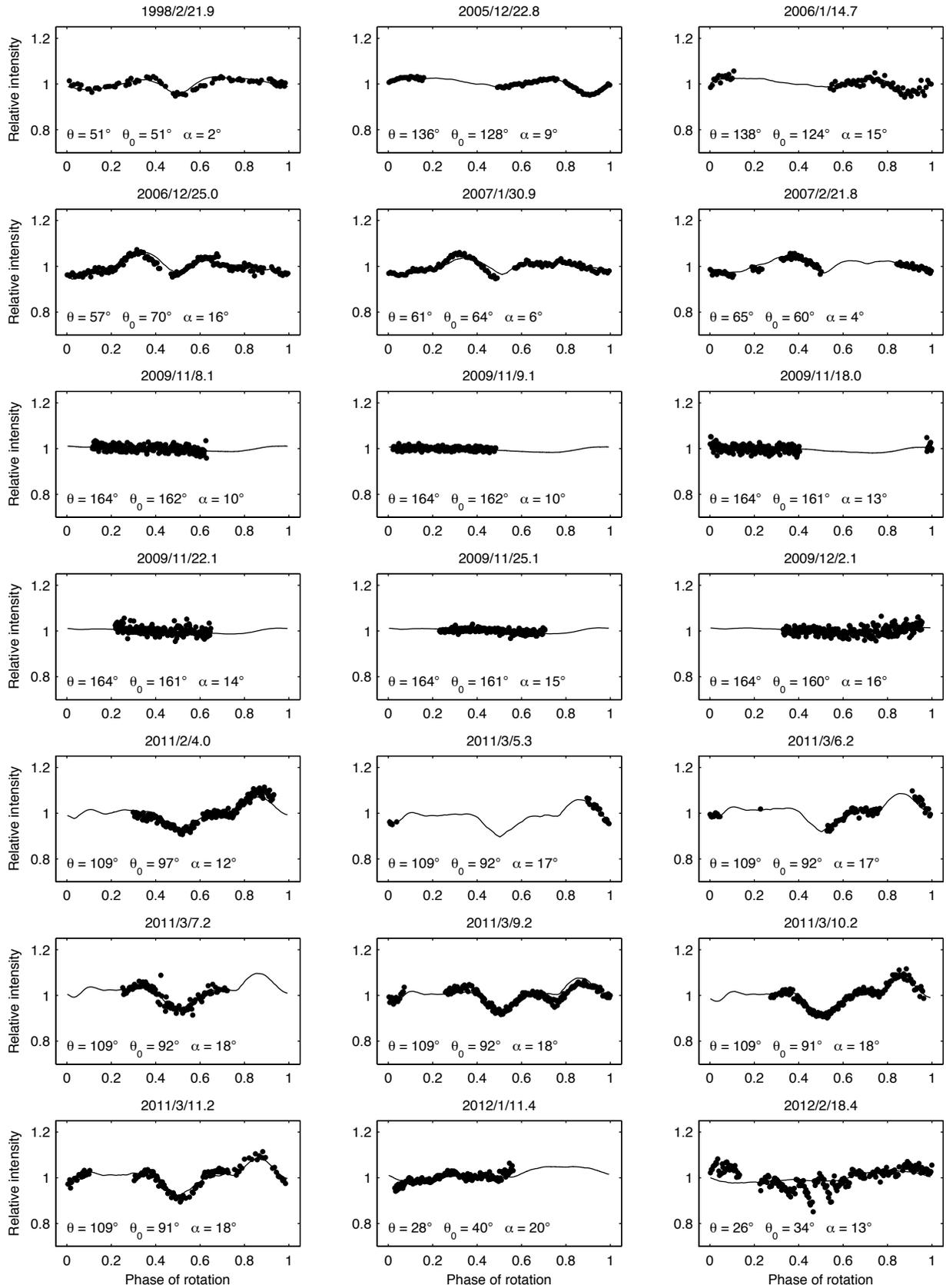

**Figure 4:** Stellar occultation chords and projected silhouette of the non-convex shape model (left and middle images) of Minerva's primary. The image on the right shows that the stellar occultation chords poorly fit the convex model derived from lightcurves only.

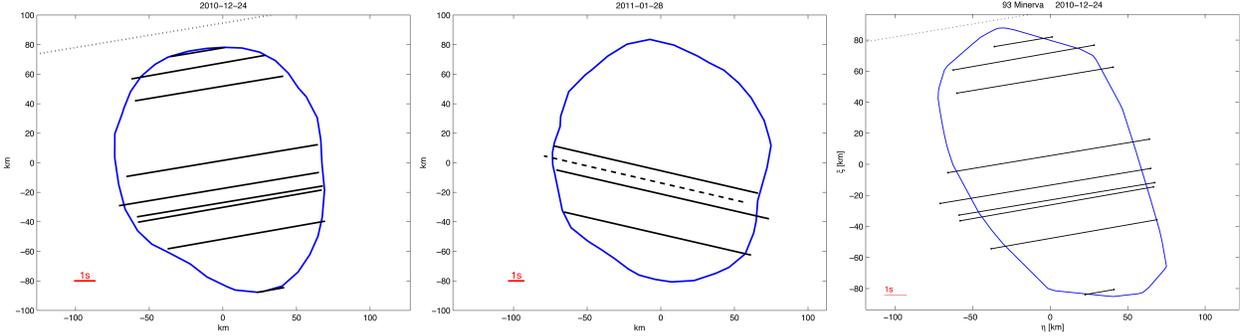

**Figure 5:** Comparison between the profile of the AO observations of Minerva's primary (red line) and the non-convex shape model (blue line). The average rms error is 3.5 *mas*, less than the 1-sigma error (4 *mas*) estimated by fitting the AO observations.

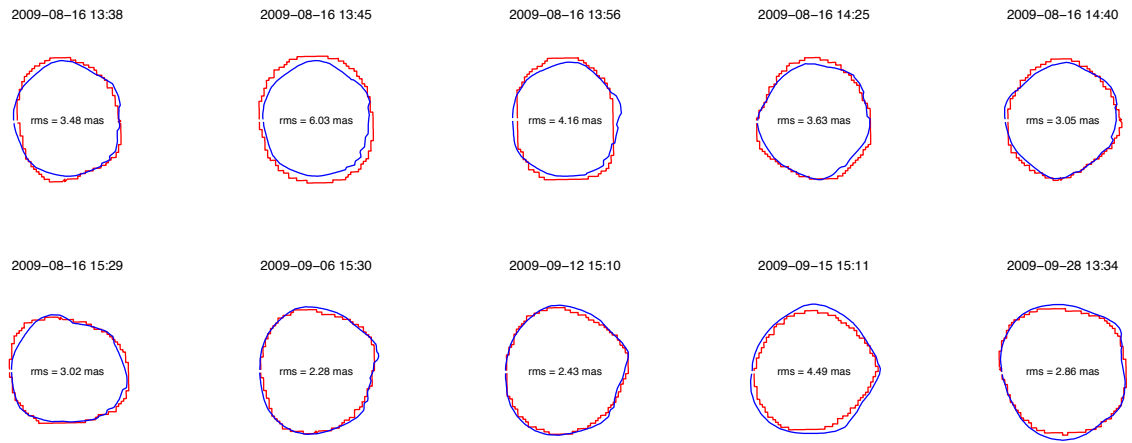

**Figure 6:** Non-convex shape model of Minerva's primary derived from combining lightcurve, adaptive optics and stellar occultation data shown from equatorial level (left, center) and pole-on (right). With a rotation period of 5.981767 ± 0.000004 h and a spin pole solution (λ=21 deg, β=21 deg in ECJ2000) with an uncertainty estimated to ± 10 deg, Minerva primary has a remarkably regular shape, leading to a quadrupole moment $J_2 \sim 0.08$ assuming an homogenous interior.

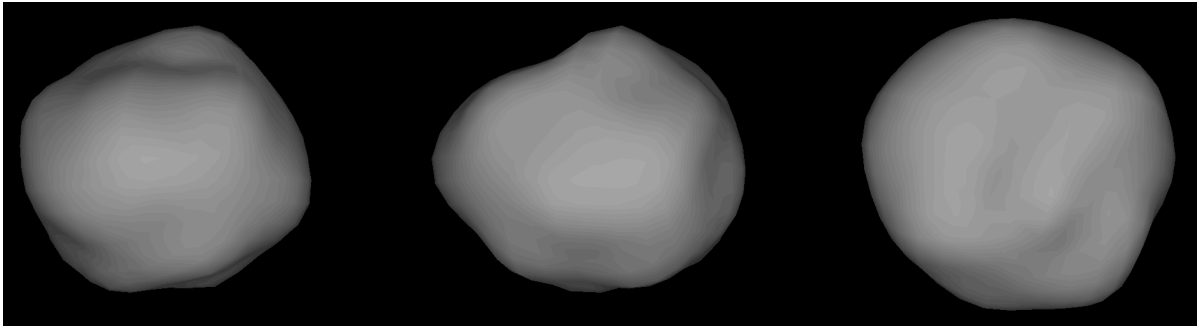

**Figure 7:** Three model fits (NEATM, STM, FRM) to the fluxes from the IRAS observation at 16:02 UT on November 6, 1983. The NEATM effective diameter ($D_{eff}$ = 154 km) and the conservative error of 15 km is in agreement with the equivalent diameter derived from our non-convex model ($D_{eq}$ = 154 ± 6 km).

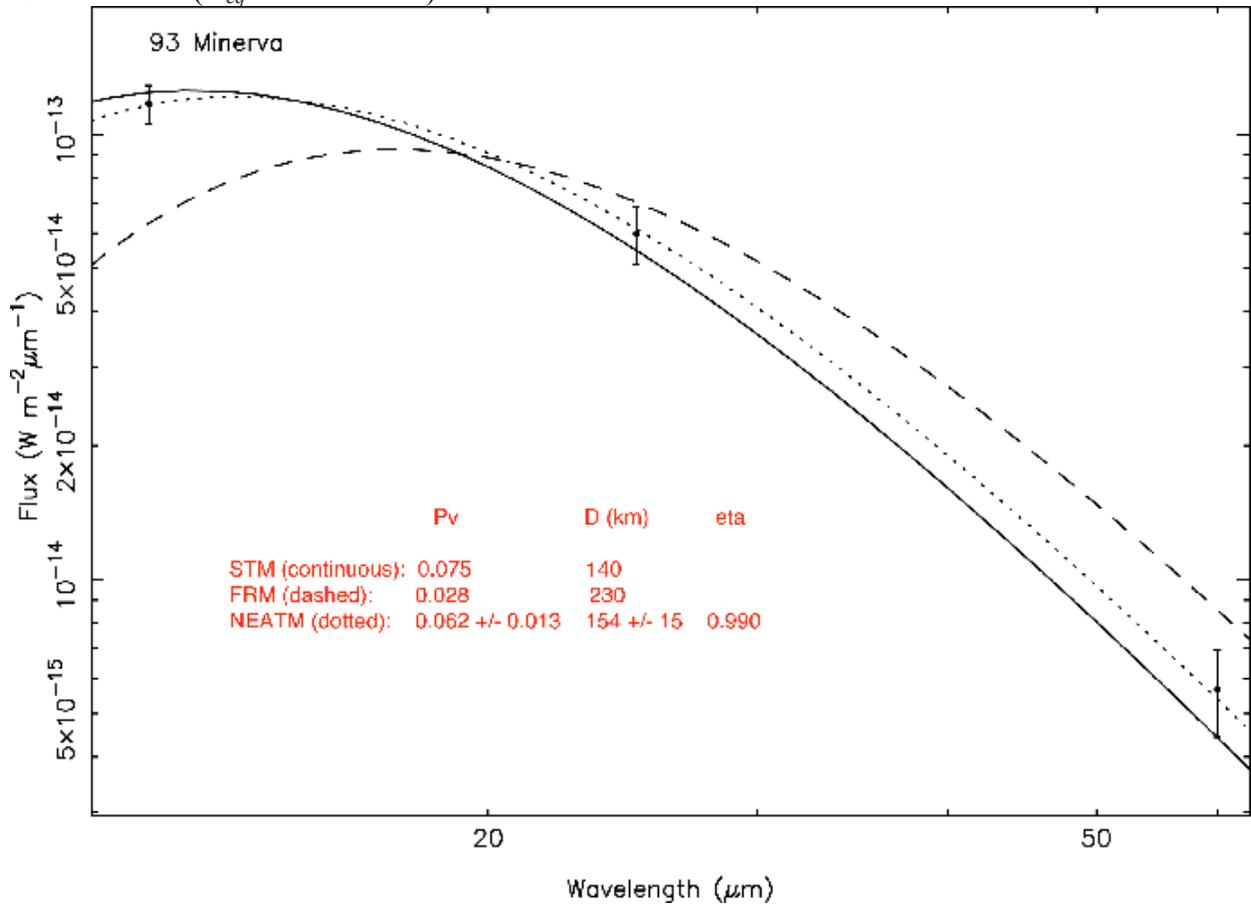

**Figure 8:** Architecture of several triple asteroid systems known in the main-belt from various sources listed in Section 4.2. In the left panel, the satellite's major axis as a function of the Hill sphere radius is represented. The right panel also displays this orbital parameter but as a function of the primary radius. (93) Minerva shares common characteristics with (45) Eugenia, (87) Sylvia and (216) Kleopatra triple asteroid systems. They are all made of a large ($D_p$ > 130 km) primary surrounded with small satellites ($D_s$ = 3 - 18 km) orbiting well inside the Hill sphere ($a$ = 1-3% × $R_{Hill}$) and close to the primary ($a$ = 5 - 12 x $R_p$). (3749) Balam is a unique triple asteroid system whose orbital parameters are not yet fully derived and its origin is likely different from the other systems.

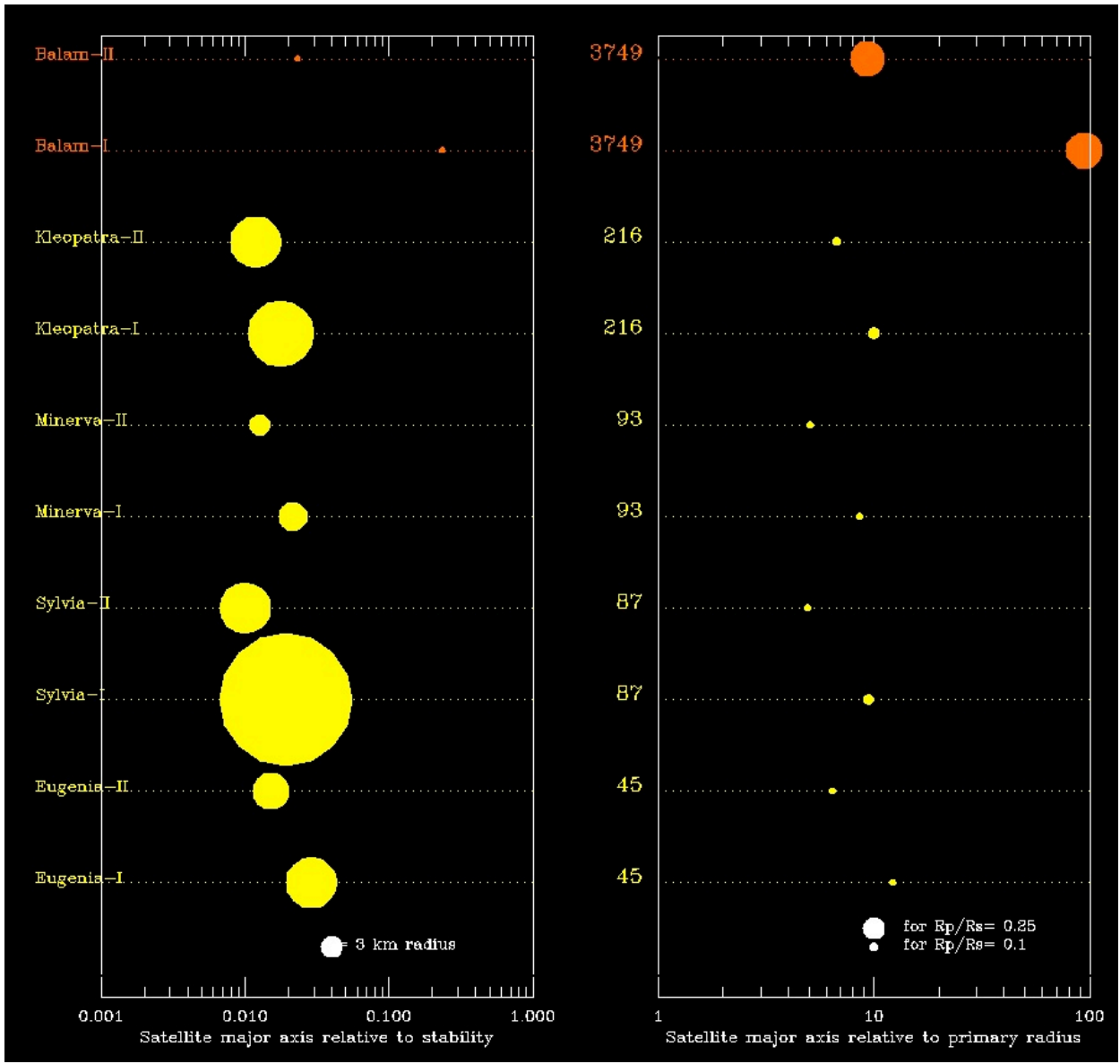

**Figure 9:** Reflectance spectra of (93) Minerva system made of a visible part from MASSII 2 stitched with a near-infrared part collected at TNG on November 1, 2009 at 00:47 UT and IRTF on September 13, 2010 at 15:19 UT and normalized at 0.55 μm. The appearance of Minerva's primary at the time of these observations is shown in the inset. The observing conditions are labeled as well (Sub-Solar-Point (SSP) and Sub-Earth Point (SEP) in planetary coordinates). The difference in latitude of 63 deg confirms that two different areas of Minerva's primary were observed. The overall shape of the two spectra is identical (weak absorption band at 0.55 μm, featureless and slightly red longward of 0.55 μm) suggesting that Minerva primary is homogeneous. By comparison with the deMeo et al. (2009) taxonomy and also considering its low albedo ($p_v \sim 0.062 \pm 0.013$), we conclude that this asteroid is a C-type.

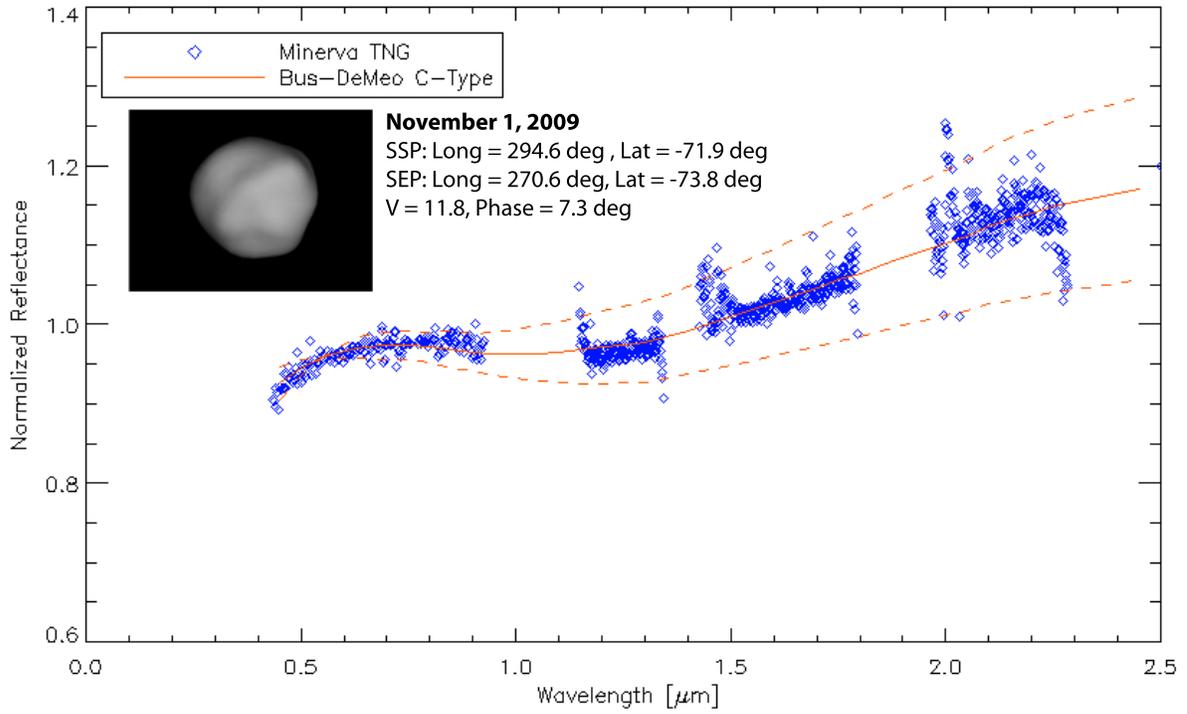

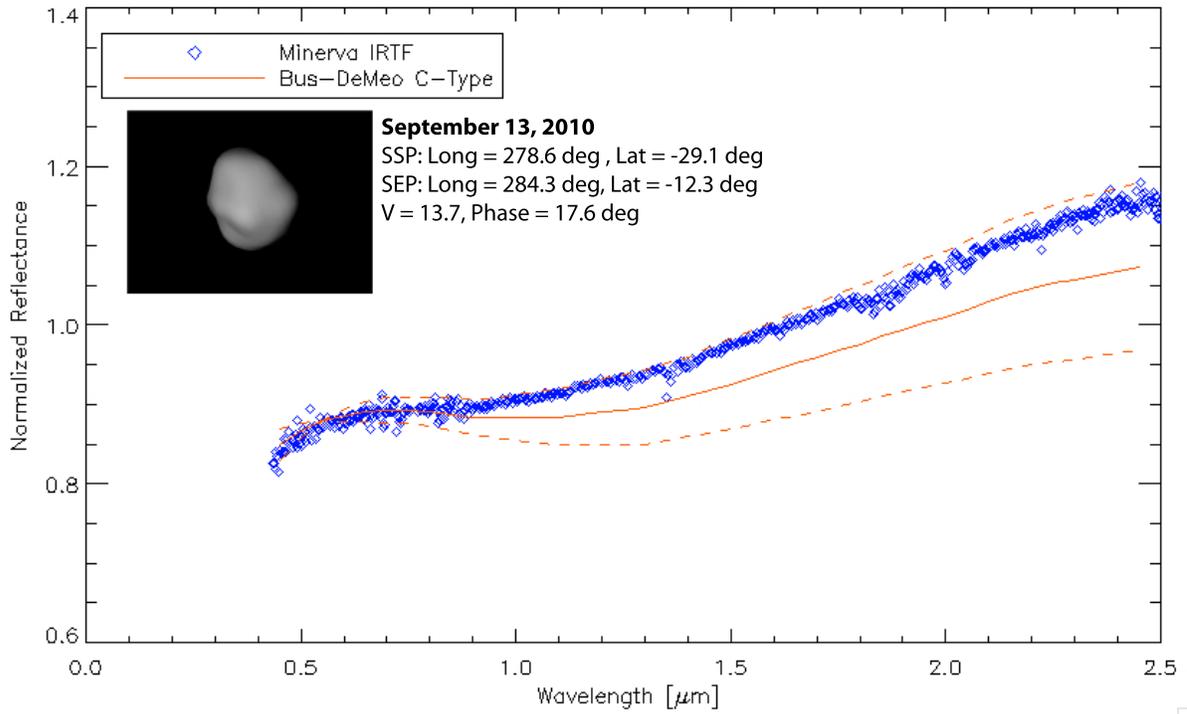

**Supplement electronic data:**
**S1:** Result of the Genoid-ANIS analysis for Minerva I
(*Minerva-I_GenoideKepler.pdf* file)
**S2:** Result of the Genoid-ANIS analysis for Minerva II
(*Minerva-II_GenoideKepler.pdf* file)
**S3:** Bus and Binzel (2002a) visible spectrum stitched with our near-infrared spectrum of (93) Minerva recorded at IRTF on September 13, 2010.
(*Minerva.vis_IR.txt*   file   ascii table)

```
Object: Minerva

Visible + NIR spectra. Stitched with "spec_stitch.pro".
More info in "~/RESEARCH/IRTF/Programs/stitches.info.txt"

  Wavelenght[um]      N.Reflec []        err [ ]
  -------------      ----------       ------------

     0.435000          0.815182          0.0411000
     0.437500          0.813978          0.0369000
     0.440000          0.808047          0.0340000
     0.442500          0.827990          0.0302000
     0.445000          0.823778          0.0270000
     0.447500          0.803577          0.0251000
     0.450000          0.830053          0.0228000
```